\documentclass[12pt]{article}
\usepackage[utf8]{inputenc}\usepackage[english,russian]{babel}
\usepackage{graphicx}

\usepackage{amsmath}
\usepackage{axodraw}
\newcommand{\be}{\begin{equation}}
\newcommand{\ee}{\end{equation}}
\newcommand{\ben}{\begin{equation*}}
\newcommand{\een}{\end{equation*}}
\newcommand{\bea}{\begin{eqnarray}}
\newcommand{\eea}{\end{eqnarray}}
\newcommand{\ar}{\begin{array}}
\newcommand{\arn}{\end{array}}

\newcommand{\vk}{\mbox{\boldmath $k$}}

\newcommand{\vks}{\mbox{\boldmath $k$}^{\;2}}

\newcommand{\q}{\mbox{\boldmath $q$}}
\newcommand{\qs}{\mbox{\boldmath $q$}^{\;2}}
\newcommand{\qp}{\mbox{\boldmath $q$}^{\;\prime}}
\newcommand{\qps}{\mbox{\boldmath $q$}^{\;\prime\; 2}}

\newcommand{\vl}{\mbox{\boldmath $l$}}
\newcommand{\vls}{\mbox{\boldmath $l$}^{\;2}}

\def\pnot{\mbox{${\not{\hbox{\kern-3.0pt$p$}}}$}}
\def\qnot{\mbox{${\not{\hbox{\kern-2.0pt$q$}}}$}}
\def\enot{\mbox{${\not{\hbox{\kern-2.0pt$e$}}}$}}
\def\knot{\mbox{${\not{\hbox{\kern-2.0pt$k$}}}$}}

\def\fun#1#2{\lower3.6pt\vbox{\baselineskip0pt\lineskip.9pt\ialign
{$\mathsurround=0pt#1\hfil##\hfil$\crcr#2\crcr\sim\crcr}}}

\makeatletter
\def\appendix{\par\clearpage
  \setcounter{section}{0}
  \setcounter{subsection}{0}
  \@addtoreset{equation}{section}
  \def\@sectname{Appendix~}
  \def\theequation{\thesection.\arabic{equation}}
  \def\thesection{\Alph{section}}}
\makeatother
\begin{document}

\begin{center}
{\large\bf   Peculiarities of Regge cuts in QCD~\footnote{Talk given at the Conference ``Hadron Structure and Fundamental Interactions:
	from Low to High Energies",  Gatchina, Russia, July 8 – 12, 2024.} }\\

\vspace{1cm}

{V.S. Fadin $^{1,2}$}

$^1${Budker Institute of Nuclear Physics of SD RAN, 
	 630090 Novosibirsk, Russia}

$^2${Novosibirsk State University, 630090 Novosibirsk, Russia}

\end{center}
\vspace{1mm}

\vspace{1mm}

\centerline{\bf  Abstract}

As is known from the classical theory of complex angular momenta, along with the Regge poles  in the complex $j$ - plane should be cuts that are generated by the poles. In  QCD,  the cuts are  generated by exchanges of the Reggeized  gluons, whose properties are strikingly different from the properties of Reggeons in classical theory, which leads  to peculiarities of the formation of the cuts. The talk is devoted to the discussion  of these peculiarities.

\vfill
E-mail:  fadin@inp.nsk.su

\newpage
\label{sec:intro}
\section{Introduction}
First of all, I would like to say a few words about the organizer of this series of conferences, Lev Nikolaevich Lipatov.  In quantum chromodynamics (QCD) there are two remarkable equations of evolution, and both of them are associated with his name. 

The first of the equations  appeared in  \cite{Lipatov:1974qm} 
as the  reformulation of  the results obtained by 
V.N. Gribov and L.N. Lipatov in \cite{Gribov:1972ri}, \cite{Gribov:1972rt}
in their investigation of deep inelastic $e^-p^+$ scattering and $e^-e^+$  annihilation 
in model field theories on the parton language. This is the equation for  evolution with the transverse momentum $Q$ of the distribution $f^a_h(x,Q^2)$ of partons $a$ carrying the fraction $x$  of the longitudinal momentum of the hadron $h$: 
\[
\frac{df^a_h(x,Q^2)}{d\ln Q^2} =\sum_b\int_x^1\frac{dz}{z}\rho^a_b(\frac{x}{z},Q^2)f^b_h f(z,Q^2)~. 
\] 
Essentially,  the equation is the renormalization group equation and its form is   universal.  It is only the kernels (splitting functions) $\rho^a_b(\frac{x}{z},Q^2)$ of the equation that depends on the form of the theory. These kernels were computed in QCD by by Y. L. Dokshitzer \cite{Dokshitzer:1977sg}  and independently G. Altarelli and G. Parisi \cite{Altarelli:1977zs}, so that the equations are now referred to as DGLAP.

The second - the equation of evolution of the gluon distribution with the fraction of the longitudinal momentum $x$ - was firstly derived  by  E. A. Kuraev,  L. N. Lipatov and me \cite{Fadin:1975cb} in  non-Abelian theories with the Higgs mechanism of mass occurrence to eliminate infrared singularities. Subsequently, its applicability was shown in QCD by I. I. Balitsky and L. N. Lipatov \cite{Balitsky:1978ic}, so that   it acquired the name BFKL. 

QCD is a unique theory in which all its elementary particles (both quarks and gluons) are Reggeized in perturbation theory. The Reggeization of an elementary particle   means that the analytical continuation $f(j, t)$ of the $t$-channel partial wave $f_j(t)$ from physical values of $j$ to the complex angular momentum plane ($j$-plane), given by the Gribov-Froissart formula \cite{Gribov:2003nw},  coincides at physical $j$ with  $f_j(t)$, despite being present in it the term with Kronecker delta-symbol due to   exchange of this elementary  particle. The gluon Reggeization is extremely important for  theoretical description of QCD processes in the Regge kinematics (high energy   $\sqrt s$ and limited momentum transfer  $\sqrt{-t}$). It provides also the factorized form of  production amplitudes in the  multi-Regge kinematics, where  all particles in the final state have limited(not growing with $s$) transverse momenta and are combined into jets with a limited invariant mass of each jet and large (growing with $s$) invariant masses of any pair
of jets.  In particular, it is the basis  of the BFKL equation.  

The gluon Reggeization provides a simple derivation of the BFKL equation not only in  the leading logarithmic approximation (LLA), when in the radiation corrections in each order of perturbation theory only the leading  powers of $\ln s$ are retained,  but in the next-to-leading logarithmic approximation (NLLA)  as well, because due to the Reggeization    amplitudes of QCD processes in Regge and multi-Regge kinematics with the adjoint  representation of the colour group in cross-channels, which are used in its derivation with the help of the unitarity relations,  are determined by the gluon Regge pole and have a simple factorized form (see \cite{Ioffe:2010zz} and  links in it).

It is known from the classical   theory of complex angular momenta \cite{Gribov:2003nw}, \cite{Collins:1977jy},  along with moving with $t$ Regge poles (Reggeons), there must be cuts in the $j$ -plane, which are commonly referred to as Regge cuts.  It is known also  that the cuts generated by the poles are also moving. 

In amplitudes with positive signature (symmetry with respect to the replacement $s\leftrightarrow u \simeq -s$), in which the real parts of the main logarithmic terms cancel, Regge cuts appear already in the LLA (in particular, the BFKL Pomeron is a two-Reggeon cut). These cuts  appear as the results of solution of the BFKL equation,  that has the strongest foundation in this approximation. Therefore, there are no particular problems with  study of these cuts. 

Problems appear in the next-to-next-to-leading logarithmic approximation (NNLLA), when the cuts appear in amplitudes with negative signature. These cuts can not be analysed using the BFKL equation, because  the amplitudes with negative signature are used themselves for the derivation of this  equation.   Due to signature conservation law  Regge cuts in amplitudes with negative signature must be at least three-Reggeon and can appear only in the NNLLA.

Regge  cuts violate the pole factorization of amplitudes 
\be
{{\cal A}}^{g^\prime g^\prime}_{gg} {{\cal A}}^{q^\prime q^\prime}_{qq} = \left({{\cal A}}^{g^\prime q^\prime}_{gq}\right)^2~, \label{pole factorization}
\ee
so such a violation may indicate the appearance of the cut contributions. 
The first observation of the violation of the Regge pole  factorization was made \cite{DelDuca:2001gu} when considering  the  two-loop amplitudes for parton ($gg, gq$ and $qq$) elastic scattering. It was observed in the  non-logarithmic (not containing $\ln s$) terms  corresponding the two-loop amplitudes to the NNLLA. 

It turned out that  the terms violating the pole Regge form  in the two-loop approximation are infrared singular. In the dimensional regularization with the space-time dimension  $D=4-2\epsilon$, they have a pole of second order in $\epsilon$. This means that the   the terms violating the  Regge pole form   can be studied using  the infrared factorization techniques. 

It was done \cite{DelDuca:2013ara}, \cite{DelDuca:2014cya} for two- and three-loop amplitudes of parton scattering. The  non-logarithmic double-pole contribution  received  at 
two-loops in \cite{DelDuca:2001gu} was confirmed  and all non-factorizing single-logarithmic infrared singular contributions at  three loops were found. 


It was natural to explain the observed violation by Regge cut contributions. 
The first explanation was done in  \cite{Fadin:2016wso}.  It is based on the consideration of Feynman diagrams, and we call it diagrammatic approach. This approach was described in more details in \cite{Fadin:2017nka}. 

Note that in contrast to the Reggeized gluon,  which contributes only to amplitudes with  the   adjoint representation of the colour group  (colour octet in QCD)  in the $t$-channel,  the  Regge cuts can contribute  to  various representations. The Regge cut  contributions for all possible colour states in two and three loops were calculated in the diagrammatic approach in \cite{Fadin:2017hnr},   and in four loops  \cite{Fadin:2019tdt} - \cite{Fadin:2021csi}.  

Another approach to calculation of  Regge cut contributions was used in  \cite{Caron-Huot:2017fxr}. 
It is  based on representation of scattering amplitudes by Wilson lines and using the shock wave approximation, and we call it  Wilson line approach. The cut contributions in two and three loops were calculated  and  the observed violation of  the pole Regge form was explained. However, the explanation differs from given in \cite{Fadin:2016wso}. The  main difference between two approaches  is the consideration of the colour structure of the cuts. It was subsequently noticed that the colour structure of the cuts assumed in  \cite{Caron-Huot:2017fxr}  disagrees with the limit of large colour number $N_c$, what was paid attention to in the papers \cite{Fadin:2019tdt} - \cite{Fadin:2021csi}. 

This drawback was eliminated in the papers \cite{Falcioni:2020lvv} - \cite{Milloy:2022fns}, where  a general recipe for separating pole and cut contributions was proposed and explicit calculations up to four loops were performed. However, the proposed recipe does not have any serious justification and, in our opinion, is not true. 
  
Thus, the properties of the three-Reggeon cuts are poorly understood. 

It must be said that also in the classical  theory of complex angular momentum, the properties of Regge cuts were studied much less well than the properties of the poles. 
Unlike the well-known criteria for the Reggeization of elementary particles in the perturbation theory \cite{Gell-Mann:1963zxa} -  \cite{Mandelstam:1964kra}  there is no criteria  for determining the cut.
But in QCD it is impossible to use even those few results that were obtained in the classical theory. There are at least two reasons for this. 
First,  in QCD all studies are based on perturbation theory, whereas in the classical  theory  the main subject of investigation  was position and nature of the $j$-plane singularities which is  not related to any fixed order of perturbation theory.  Second, unlike the classical theory, where the most important Reggeon was Pomeron with vacuum quantum numbers, having no relation to any elementary particle,  in perturbative QCD the basic Reggeon is the Reggeized  gluon possessing colour.

\section{Regge cuts in the classical theory}	
 Before appearance of QCD, the most effective tool for studying strong interactions at  high energies  $\sqrt s$  was the theory of complex angular momenta $j$ \cite{Gribov:2003nw}, \cite{Collins:1977jy}.  Significant successes in the systematization of hadrons  and in the description of the processes of their interactions  at large $s$ were achieved assuming that  the only singularities  of the partial scattering amplitudes in the complex $j$ -plane are  poles (Regge  poles, or Reggeons) moving with the square of the transferred momentum  $t$. However, it was soon realized that this assumption contradicts both   experiment,  where deviations  from the pole  factorization (\ref{pole factorization}) were  observed,  and theory, which   inconsistency  with only pole singularities  was proved. It was shown that along with the Regge  poles in a complex $ j $ - plane there should also be  cuts.

The tools of complex angular momenta turned out to be useful in the field theory of elementary particles. In papers \cite{Gell-Mann:1963zxa} - \cite{Mandelstam:1964kra},  the concept of Reggeization  of elementary particles was introduced and it was proved  that   electron in Quantum Electrodynamic (QED) with massive photon do Reggeized.   It was shown \cite{Mandelstam:1964tvk}, \cite{Mandelstam:1965zz} that it is not so, however, for  photon, so in  amplitudes  with photon exchanges  there is no Regge pole. It was also shown \cite{Gribov:1970ik}, \cite{Cheng:1970xm} that the high energy asymptotic  of  the light-light scattering amplitude   is determined by a fixed branch point that is not related to moving with $t$ cuts generated by Reggeon  exchanges.

A significant difference between Regge cuts in QCD and the classical theory of complex angular momenta is related to different ideas about Regge poles. In the classical  theory, they are associated with  infinite series of ladder diagrams Fig.1.

\begin{minipage}{7cm}
	\includegraphics[trim=0cm 20cm 0cm 0cm, clip, width=0.9\textwidth]{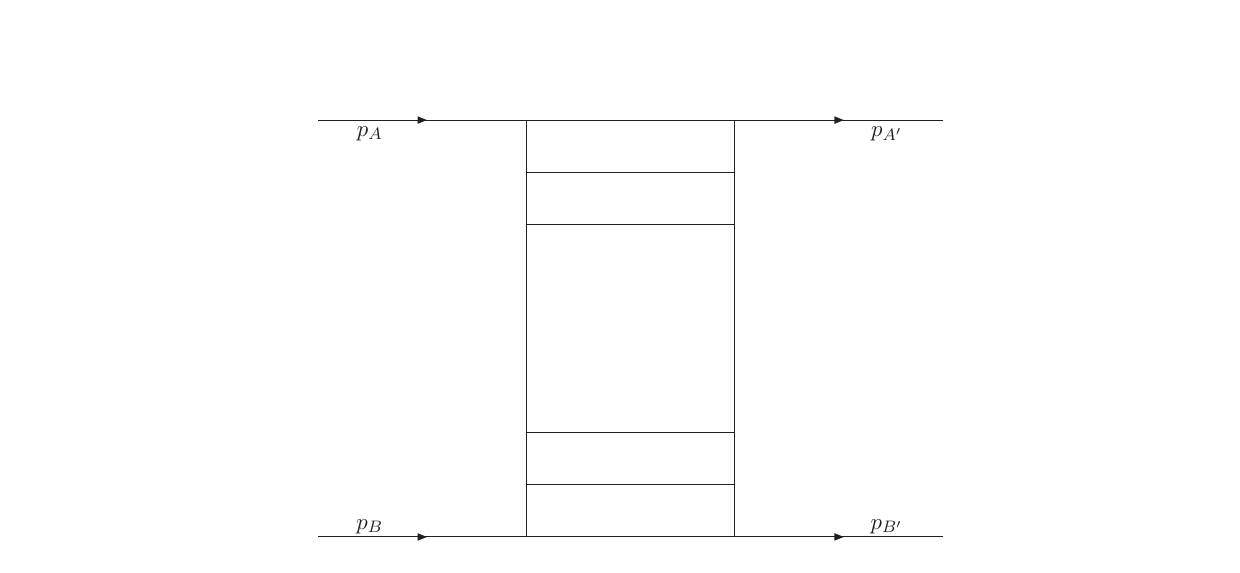}
	\label{ladder pole}
	
	\it{Fig. 1. Feynman diagrams for \\ Regge pole in the classical theory.}
\end{minipage}
\begin{minipage}{9cm}  
Such series give scattering amplitudes with asymptotic   
$A(s, t)\sim s^{\alpha(t)}$ corresponding to the contribution of  the Regge pole with the  trajectory $j=\alpha(t)$. 
It's easy to see  that the two-particle intermediate state in the  $s$-channel  for the amplitudes with two   Reggeons  with trajectories 
\be
\alpha(t) =\alpha( 0) +\alpha'( 0)t~, \label{alpha(t)}
\ee
in the $t$-channel gives the discontinuity 
\end{minipage} 

\vspace{5mm}
\noindent
	
\be
	\Delta_s A(s,t)\sim \frac{s^{2\alpha( 0) -1 -\alpha'(0) t/2}}{2\alpha'( 0) \ln s}~, \label{Delta A} 
\ee
which corresponds to moving with $t$ square root branch point at 
\be
	j_c(t)= 2\alpha( 0) -1 -\alpha'(0) t/2~. \label{j(t)}
\ee
Using this fact and considering two-particle intermediate state in the  $s$-channel unitarity relation for the amplitude corresponding to the diagram Fig.2,   D. Amati, S. Fubini and A. Stanghellini \cite{Amati:1962ey}, \cite{Amati:1962nv} came to the conclusion about the existence of the Regge cut with the branch point moving with $t$ and  passing through $2\alpha( 0) -1$ for $t=0$. Such cut and corresponding diagrams Fig.2 are called AFS cuts and AFS diagrams. 
	
\begin{minipage}{8cm} 	
	\includegraphics[trim=0cm 20cm 0cm 0cm, clip, width=0.9\textwidth]{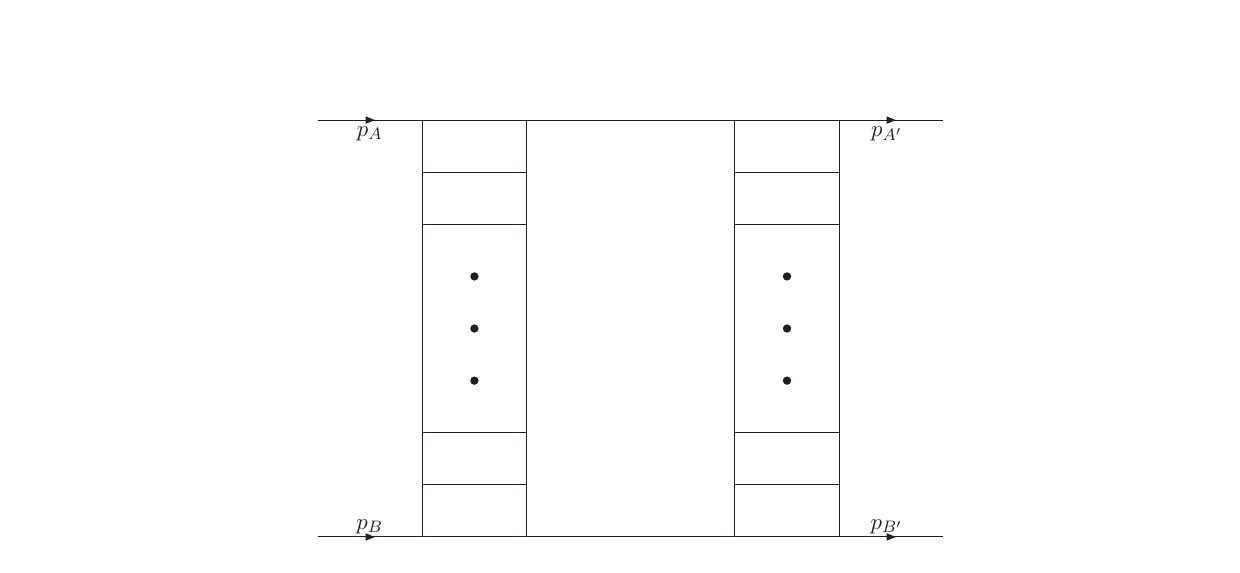}
	
	\it{Fig.2. Diagrammatic representation\\ of the AFS cut.}
\end{minipage}
\begin{minipage}{8cm} 
It is worth noting that D. Amati, S. Fubini and A. Stanghellini considered their consideration not as proof of the existence of the cut, but only as an argument in its favour. This argument was soon called into question and the mechanism they proposed for formation of cuts was soon criticized.
J.~C. Polkinghorne \cite{Polkinghorne:1963dlj} 
drew attention to the possibility of cancellation of the two-particle cuts  by another   unitarity cuts, such as  three-particle cuts given in Fig.3.  
\end{minipage}

\begin{center}
\begin{figure}[h]
		\hspace{2cm}
	\includegraphics[trim=0cm 20cm 0cm 0cm, clip, width=0.5\textwidth]{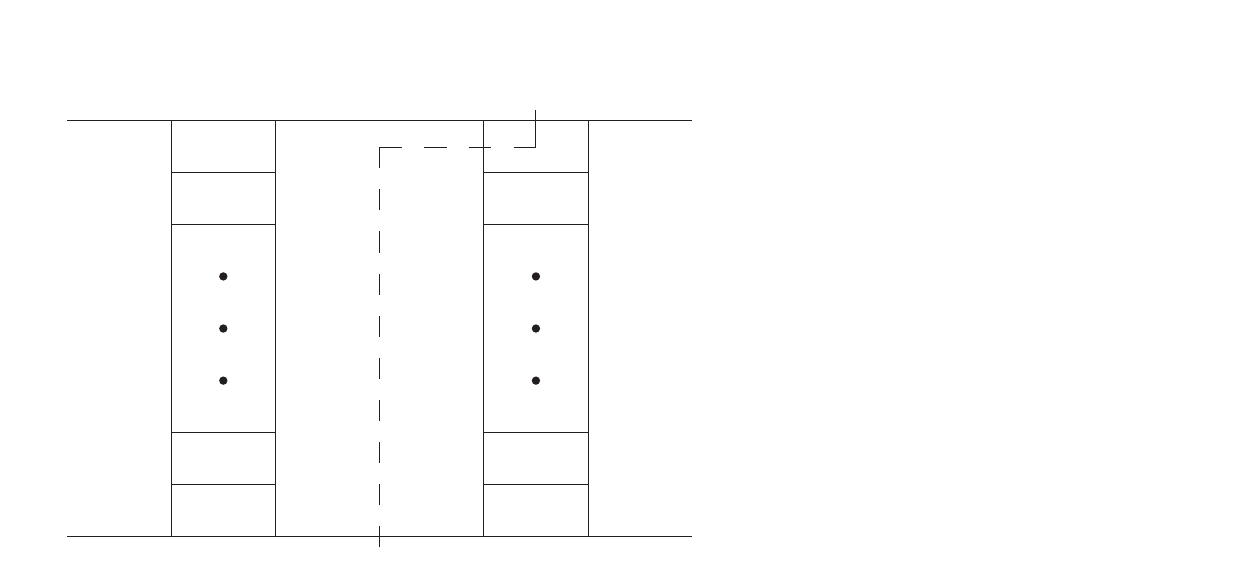}
	\includegraphics[trim=0cm 20cm 0cm 0cm, clip, width=0.5\textwidth]{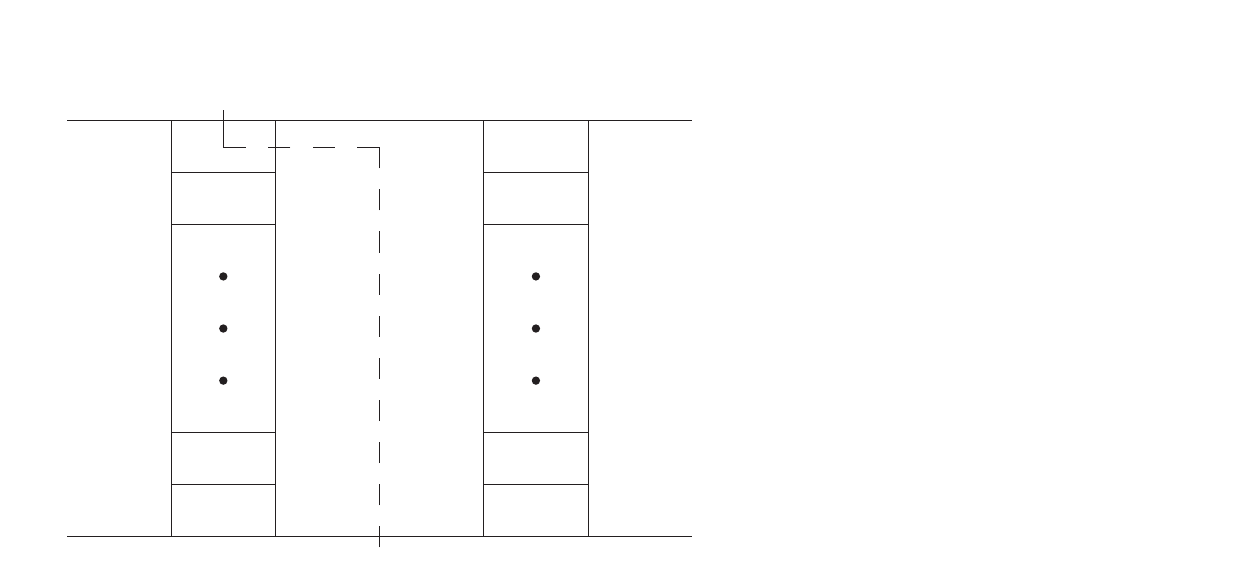}\\
	\label{fig:2}
\end{figure} 


\hspace{2cm}{\it{Fig.3. Diagrammatic representation of the three-particle \\contributions to the discontinuity of the AFS diagrams.}}
\end{center}
\vspace{0.5cm}
\noindent
Subsequently mutual cancellation of contributions of all possible unitary  cuts  and therefore  absence of a Regge cut in the total contribution of the AFS diagram was proved  in \cite{Halliday:1973vd}.
A similar conclusion was made by 
S. Mandelstam \cite{Mandelstam:1963} based on $t$-channel unitarity. 	

The cancellation of contributions of various unitary cuts is explained by the planar structure of the AFS diagram.   Denoting the amplitude as $A_1(s_1, t_1, t_2)$ where $s_1$ is  the squared invariant mass   and  $t_1$ and $t_2$ are the squared Reggeon momenta, it is easy to obtain, that the total contribution  of the AFS diagram is proportional to the integral
\be
\int_{-\infty}^{+\infty}ds_1A_1(s_1, t_1, t_2)~,  \label{nullification}
\ee
which goes to zero due to the planar structure and   decrease of  the   particle-Reggeon scattering amplitude  in  the classical theory with  increase of   the particle-Reggeon invariant mass. 
Indeed,  the integration contour should lie below all singularities for negative $s_1$ and above for positive. But due to decrease of $A_1(s_1, t_1, t_2)$ at large $s_1$ one can rotate the right side of the contour in the upper half-plane to the negative half-axis,  and to obtain zero because   there are no singularities at negative $s_1$,
due to planarity of the diagram. 

\begin{minipage}{8cm} 
	\includegraphics[width=0.9\textwidth]{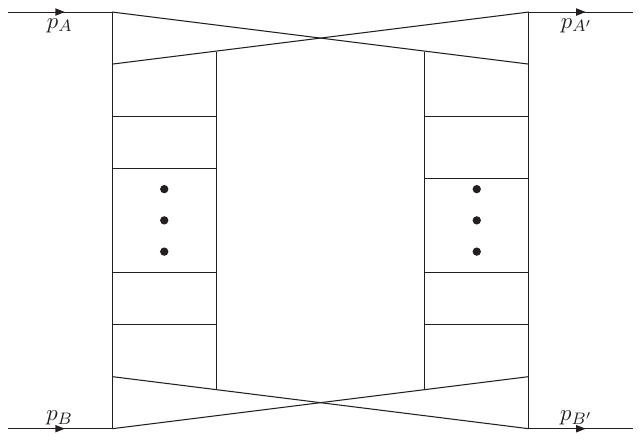}
	\vspace{-6.5cm}
	
	\it{Fig.4. Diagrammatic representation\\ of the  Mandelstam cut.}
\end{minipage}
\begin{minipage}{8cm}
To avoid such nullification
S. Mandelstam \cite{Mandelstam:1963cw}
suggested  more complicated diagrams,  like on Fig.4,  as a source of Regge cuts.  The essential point here is that these diagrams are not planar, so that Reggeon-particle scattering amplitude have both right and left cuts.

Since then, AFS type diagrams have been rejected and it was believed that only non-planar diagrams can lead to Regge cuts. 
\end{minipage}

\section{$j$--plane singularities in QCD}
	
In QCD, the situation with the cuts in the  complex angular momenta plane   is completely different.	It differs in many ways, although it seems that in  QCD the problem is the same as in the  classical  theory of complex angular momenta:  to  study  asymptotic behaviour of amplitudes at high energies and fixed momentum transfers. Moreover, Feynman diagrams are used to solve this problem in both cases. 
But in QCD we calculate  the first few terms of the perturbation theory using Feynman rules and  want to present the results of the calculations  in the form of contributions of singularities in the $j$-plane. 
In contrast, in the classical theory of complex angular momentum, one was  mainly interested in a position and a degree of singularities, which have nothing to do with the finite orders of perturbation theory.

Next, Regge poles in the  classical  theory are represented by infinite series of ladder diagrams  in model field theories that have no relation to reality. Such representation is used just to illustrate some properties of Regge poles.  In particular, it can not be applied to  the main Reggeon in the  classical  theory, which  is the Pomeron with the trajectory $j_P(t)$ and positive signature. Instead, the Reggeized gluon with the  trajectory $j(t) = 1 + \omega(t)$ and negative signature, which is is the main Reggeon in QCD, is  well described by Feynman diagrams. But these diagrams  have hot ladder structure and start with one-gluon exchange.  

Besides this, the important difference is that the  Reggeized gluon carries colour, while this quantum number was absent in  the  classical  theory.

\subsection{Two-Reggeon cuts in QCD}
Regge cuts appear already  in the LLA in  amplitudes with two Reggeized  gluons  in the  $t$-channel and positive signature, where  real parts of the leading logarithmic terms cancel out, so that remaining piece is pure imagine in the LLA. In particular, the BFLK Pomeron is a two-Reggeon cut.

Using the pole Regge form of elastic and MRK amplitudes, one obtains from the unitarity relation the elastic scattering amplitudes  as the convolution

\be	
	\Phi _{A^{\prime }A}\;\otimes \;G\;\otimes \;\Phi _{B^{\prime}B}~. \label{FGF}
\ee	
\begin{minipage}{8cm} 
		\includegraphics[width=0.9\textwidth]{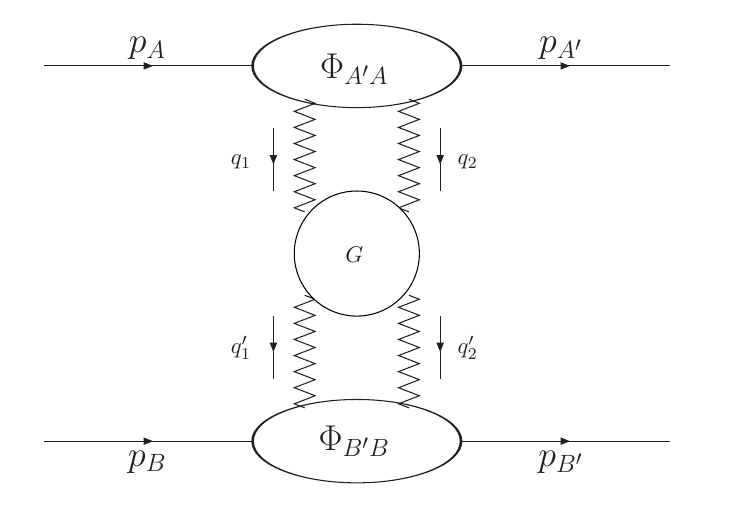}
\it{Fig.5. Diagrammatic representation\\ of two-Reggeon exchange.}
\end{minipage}
\begin{minipage}{8cm} 
Here impact factors $\Phi _{A^{\prime}A}$ and
$\Phi _{B^{\prime }B}$ describe  transitions
$A\rightarrow A^{\prime
}$ and $B\rightarrow B^{\prime }$, 
$G$ is so called   Green's function for two
interacting Reggeized gluon.  In the angular momentum space,  it is given  by the operator
\be {\hat{\cal
		G}}=\frac{1}{j-1 -\hat{\cal{K}}}~, \label{G} 
\ee
where  ${\hat{\cal{K}}}$ presents the  BFKL kernel, 
\end{minipage}
\be
 \hat{\cal {K}}=\hat{\omega}_1+\hat{\omega}_2+ {\hat{\cal {K}}_r}~, 
\ee
$\hat{\omega}_{1,2}$  present  the Reggeized gluon trajectories,		
and $\hat{\cal {K}}_r$ stands for so called real part of the BFKL kernel describing interaction of two Reggeized gluons.
In the transverse momentum space, in the leading order 
\be
\omega_i =  \omega(-\qs_i)
=-g^2 N_c {\qs_i} \int\frac{d^{2+2\epsilon}l}{2(2\pi)^{(3+2\epsilon)}\vls(\q_i -\vl)^2}~. \label{omega i} 
\ee
For Reggeons with transverse momenta $\q_1$ and $\q_2$  and colour indices $c_1$ and $c_2$.\\
\be
\left[{\cal K}_r(\q_1, \q_{2}; \vk)\right]^{c'_1c'_2}_{c_1c_2} =
-T_{c_1c'_1}^{a}T_{c_2c'_2}^{a}\frac{g^2}{(2\pi)^{D-1}}
\left[\frac{\qs_{1}\qps_{2}+\qs_{2}\qps_{1}}
{\vks}-\qs\right]~, \label{K_r}
\ee
\[
\qp_{1}= \q_1-\vk, \;\; \qp_{2}= \q_2+\vk~. 
\]
\noindent Energy dependence of
scattering amplitudes is determined by the  BFKL kernel.

The BFKL kernel and the impact factors are expressed in terms of the Reggeon vertices and trajectory. The kernel is universal (process independent). 

In the Pomeron (colour singlet and positive  signature) channel the leading singularity is  a square root branch branch point \cite{Fadin:1975cb}, \cite{Ioffe:2010zz} at  
	\be
	\omega_P=\frac{4N_c\alpha_s}{\pi}\ln 2~. \label{Pomeron intercept}
	\ee  
Thus, in the BFKL approach, two-Reggeon cuts appear already in the LLA. But their properties differ drastically from two-Reggeon cuts in the classical theory. In particular, the BFKL Pomeron is a fixed branch point (\ref{Pomeron intercept}), instead of moving branch points (\ref{j(t)}) in the classical theory. Then, it arises due to planar diagrams, which is evident from the fact that its intercept $\omega_P$ (\ref{Pomeron intercept}) is not suppressed at  large number of the colours $N_c$. It means that the Mandelstam arguments \cite{Mandelstam:1963} don't work in QCD. Evidently,   
this is due to the different nature of Reggeons. It is clear from the lowest order contribution  to amplitudes with positive signature, which is given by diagrams with two-gluon exchange.  This contribution has both $s$- and $u$-channel discontinuities, although it is given be planar  diagrams. 

It should be noted that (\ref{Pomeron intercept}) gives the position of the leading (rightmost) singularity in the colourless channel. The existence of this singularity  does not at all exclude the presence of other singularities. Generally speaking, structure of  $j$-plane singularities is rather complicated and  is different for different representations of the colour group in the $t$-channel. It was shown \cite{Kuraev:1977fs} in the  non-Abelian theories with the Higgs mechanism of mass formation that there are there also moving poles and cuts in amplitudes with positive signature.  Unfortunately, structure of $j$-plane singularities of the amplitudes (\ref{FGF}) 
in QCD is not well investigated.
	
 It may seem that the  amplitude (\ref{FGF}) contains different from the Reggeized gluon pole singularities  also in the channel with gluon quantum numbers and negative signature.
It turns out, however, that this is not the case thanks to the  bootstrap relations (see \cite{Ioffe:2010zz} and links in it). 

These relations are quite simple in the elastic case:
\begin{center}
\includegraphics[trim=0cm 20cm 0cm 0cm, clip, width=0.7\textwidth]{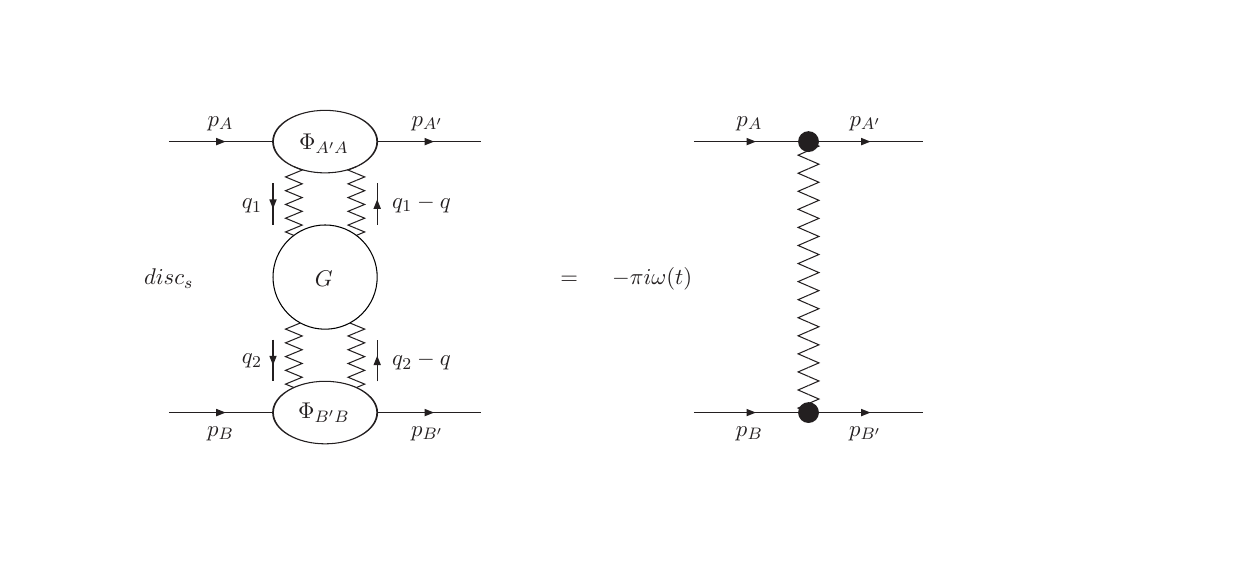}
\end{center}
\vspace{-1cm}
\hspace{2cm}{\it{Fig.6.  Diagrammatic representation  of the bootstrap relation.}}
\vspace{0.5cm}

\noindent
There is an infinite number of the bootstrap relations for  production amplitudes in the MRK, and they provide the pole Regge form not only in the LLA, but  in the NLLA as well (see \cite{Ioffe:2010zz} and links in it). 

It should be noted that   appearance of a  Reggeon with negative signature  in the  channel with  two Reggeons with the same signature  is forbidden by  the signature conservation law \cite{Gribov:2003nw}.  The point here is that Fig.5   and Eqs. (\ref{FGF}-\ref{K_r})  do not actually represent the  two-Reggeon exchange in the sense of the classical theory.  Actually they  represent  amplitudes,  reconstructed by analyticity  from  the s-channel discontinuities of elastic amplitudes, coming  in the unitarity relations   from intermediate states with Regge and multi-Regge kinematics.  It is the amplitudes  that are  called in the BFKL approach the amplitudes with the exchange of two Reggeized  gluons.

Thus, there is a huge difference between cuts considered in the classical theory and in QCD: in nature of Reggeons, in their interaction, and  in their colour. This difference makes impossible to apply experience of classical theory to perturbative QCD, especially taking into account that here we we want to identify with the cut contributions with  terms of finite orders of perturbation theory, have no relation either to the position or to the nature of the $j$-plane singularities of the  partial amplitudes.

\subsection{Three-Reggeon cuts in the diagrammatic approach}
By analogy with the two-Reggeon cut, the contribution of the three-Reggeon cut  is depicted by Fig.7 and is represented as
\be	
\Phi^{3R} _{A^{\prime }A}\;\otimes \;G^{3R}\;\otimes \;\Phi^{3R} _{B^{\prime}B}~.\label{F3G3F3}
\ee	
\begin{minipage}{8cm} 
		\includegraphics[trim=0cm 17cm 0cm 0cm, clip, width=1.1\textwidth]{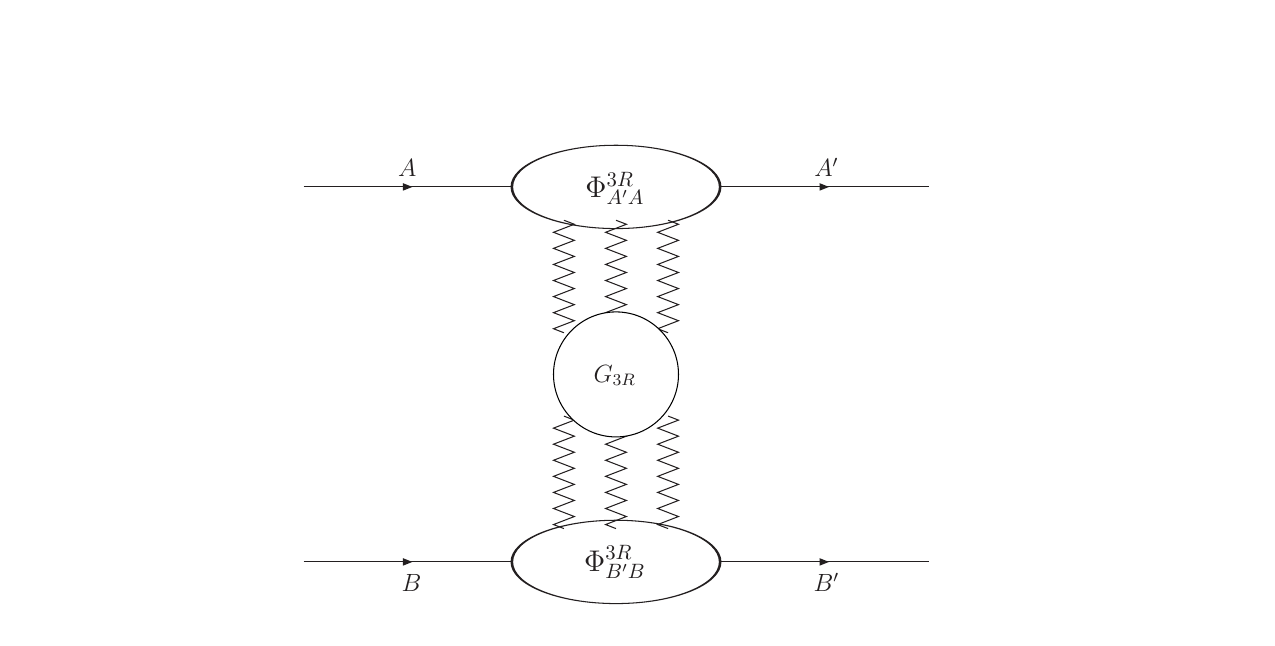}
	
	\it{Fig.7. Diagrammatic representation\\ of the three-Reggeon exchange.}
\end{minipage}
\begin{minipage}{8cm} 
\be 
{\hat{\cal G}^{3R}}=\frac{1}{\omega -{\hat{\cal{K}}_{3R\rightarrow 3R}}}~, \;\;\;\;\;\;\;\; \omega = j-1, \label{G3}
\ee

\be
{\hat{\cal{K}}_{3R\rightarrow 3R}} = {\hat{\cal
{\omega}}_1}+{\hat{\cal {\omega}}_2}+{\hat{\cal {\omega}}_3}+ {\hat{\cal {K}}}_r^{12}+ {\hat{\cal {K}}}_r^{13}+ {\hat{\cal {K}}}_r^{23}~,\label{K_r3}
\ee

$K_r^{ij}$ describes interaction of Reggeized gluons $i$ and $j$.  It is supposed that  pair interaction between Reggeons is described  by  the BFKL kernel.
In fact, such interaction was assumed  many years ago  in the BKP equation \cite{Bartels:1980pe}, \cite{Kwiecinski:1980wb} for the Odderon (colourless state of three 
\end{minipage}

\noindent
Reggeized gluons with positive signature, which differs from Pomeron by C-parity).

This assumption looks rather natural, although,  as I know, its strict proof does not exist, as well as its check in perturbation theory.

In the leading (two loop) approximation Reggeon interaction must  be neglected. The exchange of non-interacting Reggeons is depicted in  Fig.8 $a$. In the momentum space, the interaction of Reggeons with particles depends only on the total energy of the process and on the transverse momenta  of the Reggeons. Therefore in the momentum space the lowest order contribution is depicted by Fig.8$b$, where solid lines correspond to the propagators $ \frac{1}{\mathbf{k}_{i\perp^2}}$. 


\begin{center}
		\includegraphics[trim=0cm 18cm 0cm 4cm, clip, width=1.1\textwidth]{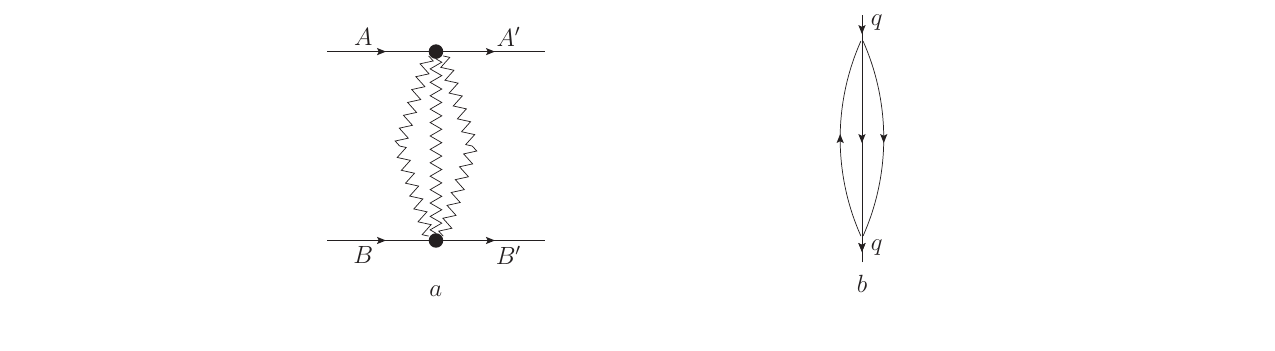}

\hspace{2cm}
{\it{Fig.8.  Diagrammatic representations  of the  two-loop contribution.}}
\end{center}
\vspace{0.5cm}

\noindent 
The colour structure of the cut  is determined using the results \cite{DelDuca:2013ara}, \cite{DelDuca:2014cya},  based  on the infrared factorization. It will be discussed in the next subsection.  For separation of pole and cut contributions, difference  in  their energy dependence was used in the three-loop approximation, in which   the cut contribution is depicted by the Reggeon diagram Fig.9$a$,  

\vspace{-0.5cm}
\begin{center}
\includegraphics[trim=0cm 20cm 0cm 4cm, clip,width=1.1\textwidth]{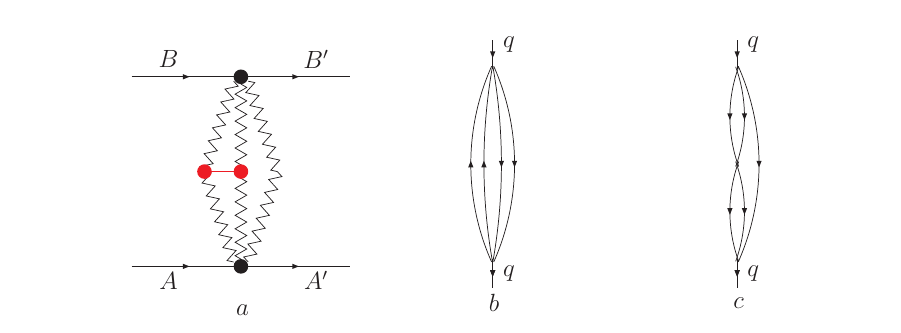}

{\it{Fig.9. Diagrammatic representation of three-loop contribution.}}
\end{center} 

\noindent 
where the horizontal line represents the BFKL kernel, and its momentum pars by two diagrams, Fig.9$b$  and  Fig.9$c$. 
Accordingly, in the four-loop approximation the Reggeon diagrams of the cut contribution are  depicted in Fig.10   and their  momentum parts in in Fig.11.

\vspace{-5mm}
\includegraphics[width=0.9\textwidth]{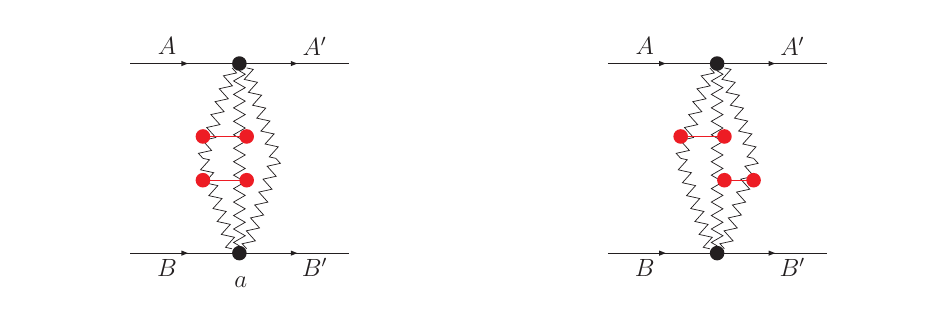}

\hspace{2.5cm}{\it{Fig.10. Reggeon diagrams for the  four-loop contribution.}}

\vspace{-5mm}

\includegraphics[width=0.9\textwidth]{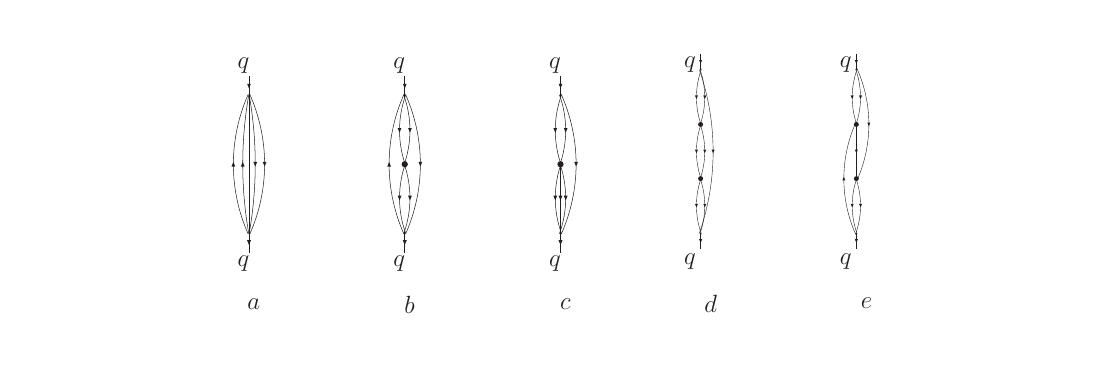}

\hspace{1cm}{\it{Fig.11.  Transverse momentum diagrams for the  four-loop contribution.}}
  
\subsection{Colour structure of the cuts} 
A crucial question is the colour structure of Reggeon cuts. 
This question is simple  in the two-Reggeon case, because in the product of two  adjoint representations there is only one representation of given dimension and parity.   But it becomes quite non-trivial in the case of three-Reggeon  cuts. 
The Reggeon cut contributions are obtained as the sums of their momentum parts with the colour coefficients. In the diagrammatic approach the colour coefficients are determined  from two- and three-loop approximations.

Due to the signature conservation the cut with  negative signature must be the three-Reggeon one. 
Since our Reggeon is the Reggeized gluon, the three-Reggeon cut first contribute to amplitudes corresponding to  the diagrams depicted in Fig.12. 

\vspace{.5cm}
\begin{center}
\includegraphics[trim=0cm 1.5cm 0cm 0cm, clip,width=1.0\textwidth]{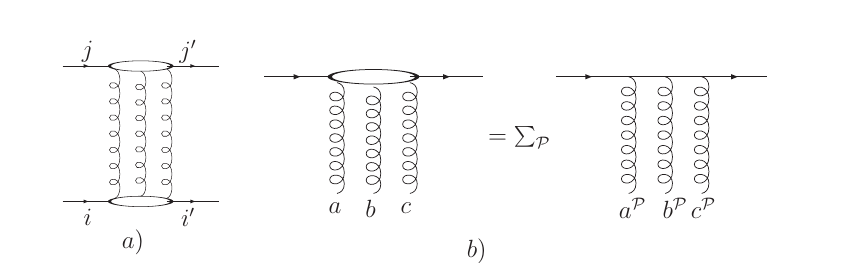}

{\it{Fig.12.  Three-gluon exchange  diagrams.}}
\end{center}
Remind that unlike  the Reggeon,  which contributes only to amplitudes with  the   adjoint representation of the colour group  (colour octet in QCD)  in the $t$-channel,  the  cut can contribute  to  various representations. Besides the octet ($\bf{8}$), possible representations   are  singlet ($\bf{1}$) for quark-quark and quark-gluon scattering and $\bf{10}$,  $\bf{10^*}$ and  $\bf{27}$ for  gluon-gluon scattering (remind that we consider only negative signature).  The colour coefficients of the diagrams Fig.12 depend on the representation.  It occurs that they are equal for all  diagrams for the representations different from the adjoint. For these representations,  we attribute all diagrams to the contribution of the three-Reggeon  cut. Note that the sum of the contributions of all diagrams with equal coefficients does not depend on the gluon gauge, which makes our definition gauge invariant.

As for the adjoint representation, the colour coefficients are equal separately for two  planar and four non-planar diagrams, but differ  for planar and non-planar diagrams.  In this representation, the contribution to the amplitude can come from both the Regge pole and the three-Reggeon cut, and the problem of separating these contributions arises. In the diagrammatic approach, this problem   is solved  using the results \cite{DelDuca:2013ara}, \cite{DelDuca:2014cya} for the infrared singular contributions  to two- and three-loop amplitudes.  The colour coefficients of the three-loop contributions differ from the one-loop coefficients only by common factors, different for  for the  pole and cut contributions.  Comparison of the obtained results with the results of \cite{DelDuca:2013ara}, \cite{DelDuca:2014cya}  based on the infrared factorisation  allows to determine the colour coefficients of the cut contribution. It occurs that they also  are equal for all  diagrams, as well as  for the representations different from the adjoint. The colour coefficients of the four-loop contribution are determined by considering that the separation into pole and cut contributions is specified by two and three loops.

In the  Wilson line approaches certain assumptions are made about the colour structure of the cuts from the beginning. Using these assumption demanded introduction of the  three-Reggeon cut-one Reggeon  mixing for the explanation of the violation of the pole factorization found in \cite{DelDuca:2001gu}-\cite{DelDuca:2014cya}. 
As it was already discussed, the assumption made in the first version  of this approach \cite{Caron-Huot:2017fxr} contradicts to the known results of N=4 super-Yang-Mills theory in the planar limit. In the modern version, used in \cite{Falcioni:2020lvv} - \cite{Milloy:2022fns}, for the separation  of pole and cut contributions  is  based on the assertion that the diagrams for Regge cuts are non-planar. As it was already discussed, this assertion, which comes from the classical  theory of complex angular momenta, is
inapplicable in QCD. In addition, separate  contributions of planar and non-planar diagrams are  not gauge invariant.

\section{Conclusion}
A remarkable property of QCD, extremely important for describing processes in the Regge and the multi-Regge kinematics, is the gluon  Reggeization. Thanks to it, the amplitudes of processes with adjoint colour group representation in cross-channels and negative signature have the pole Regge form in the LLA  and the  NLLA, which ensures a simple derivation of the BFKL equation. 

As is known from the classical theory of complex angular momenta,  Regge poles generate cuts. The $j$-plane cuts  appear in amplitudes with positive signature already in the  LLA as the results of the BFKL equation, so that their investigation has  a solid foundation. It shows that properties of these cuts  differ drastically from properties of two-Reggeon cuts in the classical theory.  In particular, the BFKL Pomeron is a fixed branch point, instead of moving branch points  in the classical theory. Then, it arises due to planar diagrams, which is evident from the fact that its intercept $\omega_P$ (\ref{Pomeron intercept}) is not suppressed at  large number of the colours $N_c$. It means that the Mandelstam arguments \cite{Mandelstam:1963} don't work in QCD. 

The situation is completely different with cuts in amplitudes with negative signature, which themselves are used to derive the BFKL equation.  At these amplitudes, Regge cuts  appear only in the NNLLA. Evidence of their appearance is the violation of the pole factorization of  elastic amplitudes factorization \cite{DelDuca:2001gu}-\cite{DelDuca:2014cya}. Unfortunately,  the experience of the classical theory of complex angular momenta cannot be used for their investigation,  as it is seen from the case of amplitudes  with positive signature, both due to the different nature of Reggeons and due to differences in the purposes and approaches to study  of $j$-plane singularities.

There are currently two approaches to explaining the violation of factorization by contributions from three-Reggeon cuts: the  diagrammatic approach, \cite{Fadin:2016wso}-\cite{Fadin:2021csi}, which is  is based on the analysis of Feynman diagrams, and the  
 Wilson line approach, \cite{Caron-Huot:2017fxr} - \cite{Milloy:2022fns},  based on using  of representation of scattering amplitudes by Wilson lines and shock wave formalism. The main difference between these two  approaches is the colour structure of the three-Reggeon cut contributions. In both approached the violation of the pole factorization found in \cite{DelDuca:2001gu}-\cite{DelDuca:2014cya} was  explained (but in different  ways)  and four-loop cut contributions are calculated for all colour group representations in the $t$-channel. In the Wilson line approach, a general recipe for separating pole and cut contributions was proposed. The  recipe proposed in the first version of the Wilson line approach \cite{Caron-Huot:2017fxr}  contradicts the maximally extended supersymmetric Yang-Mills theory in the planar limit.  In the modern version of this  approach  \cite{Falcioni:2020lvv} - \cite{Milloy:2022fns} separation  of the cut contributions as contributions of non-planar diagrams  is based on the Mandelstam arguments \cite{Mandelstam:1963}, \cite{Mandelstam:1963cw}. But these arguments are not valid in QCD, as it clearly seen in the case of the two-Reggeon cuts, precisely  for which these arguments were formulated. Besides this, separate  contributions of planar and non-planar diagrams are  not gauge invariant. In our opinion, all this makes one doubt the proposed recipe.  

Further development of  theory of  QCD processes in the Regge region call for better understanding of  the  three-Reggeon cuts. 

Standard derivation of the BFKL equation requires calculations  of contribution of these cuts not only in elastic amplitudes, but in amplitudes with the multi-Regge kinematics as well.

\end{document}